\documentstyle[12pt]{article}
\input psfig
\begin{document}
\title{PERTURBATIVE\ ANALYSIS\ OF\ CHERN-SIMONS\
FIELD\ THEORY\ IN\ THE\ COULOMB\ GAUGE}
\author{Franco Ferrari \thanks{The work of F. Ferrari
has been supported by the European Union, TMR Programme, under
grant ERB4001GT951315} $^{,a}$  and Ignazio Lazzizzera
$^{b,c}$\\ \\
$^a$ {\it LPTHE \thanks{Laboratoire associ\'e No. 280 au CNRS} ,
Universit\'e Pierre er Marie Curie--PARIS VI and} \\
{\it Universit\'e Denis Diderot--Paris VII,
Boite $126$, Tour 16, $1^{er}$ \'etage,}\\
{\it 4 place Jussieu,
F-75252 Paris CEDEX 05, FRANCE.} \\
$^b$ {\it Dipartimento di Fisica, Universit\'a di Trento, 38050 Povo (TN),
Italy.}\\
$^c$ {\it INFN, Gruppo Collegato di Trento, Italy.}
}

%\date{October 96}
\maketitle
\vspace{-4.5in} \hfill{Preprint PAR-LPTHE 96--44, UTF 387/96} \vspace{4.4in}
\begin{abstract}
In this paper we analyse the perturbative aspects of
Chern--Simons field theories in the Coulomb
gauge.
We show that in the perturbative expansion of the Green functions
there are neither
ultraviolet not infrared
divergences.
Moreover, 
all the radiative corrections are zero at any loop order. 
Some problems connected with the Coulomb gauge fixing, like the appearance
of spurious singularities in the computation of the Feynman diagrams,
are discussed and solved.
The regularization used here for the spurious singularities 
can be easily applied also to the Yang--Mills case, which is affected
by similar divergences.
\end{abstract}
\section{Introduction}

In the recent past, the Chern--Simons (C-S) field
theories \cite{jao,hageno} have been intensively studied in connection
with several
physical and mathematical applications \cite{csapps,ienkur}.
A convenient gauge fixing for these theories is provided by the Coulomb gauge.
As a matter of fact, despite of the presence of nontrivial
interactions in the gauge fixed action, the calculations become
considerably simpler than in the covariant gauges and a perturbative
approach is possible also on non-flat manifolds \cite{ffprd}.
Moreover, the dependence on the time in the Green functions is trivial,
so that the C--S field theories can be treated in practice as two
dimensional models.
Starting from the seminal works of refs. \cite{hageno,hagent} and \cite{jat}, 
the Coulomb gauge has been already applied in a certain number of
physical problems involving C--S based models
\cite{ienkur}, \cite{vo}--\cite{bcv}, but still remains
less popular than the covariant and axial gauges.
One of the main reasons is probably the fact that there are
many
perplexities concerning the use of
this gauge fixing,  in particular in the case of the four dimensional
Yang--Mills theories \cite{taylor}--\cite{leiwil}.
Recently, also the consistency of the C--S field theories in the
Coulomb gauge has been investigated using various techniques \cite{ffprd,
devone, cscgform, frig},
but so far a detailed perturbative analysis in the non-abelian case
is missing.
To fill this gap, the radiative corrections of the Green functions
are computed here at any loop order and it is shown
that they vanish identically.
No regularization is needed for the ultraviolet and infrared
divergences since, remarkably, they do not appear in the amplitudes.
The present result agrees with the previous analysis
of \cite{frig}, in which the
commutation relations between the fields are proved to be trivial
using the
Dirac's canonical approach to constrained systems.
It is important to notice that the absence of any quantum
correction despite of the presence of nontrivial self-interactions in the
Lagrangian
is a peculiarity of the Coulomb gauge that cannot be totally
expected from the fact that the theories under consideration
are topological, as finite renormalizations of the fields and
of the coupling constants are always possible.
For instance, in the analogous case of the covariant gauges,
only the perturbative
finiteness of the C--S amplitudes has been shown \cite{csformal}
in a regulatization
independent way exploiting BRST techniques \cite{ss}.
Indeed, a finite shift of the C--S coupling constant has been observed
in the Feynman gauges by various authors \cite{shift, alr}.

The material presented in this paper is divided as follows.
In Section 2 the C--S field theories with $SU(n)$ gauge group are quantized
using the BRST approach. The Coulomb gauge constraint
is weakly imposed and
the proper Coulomb gauge is recovered suitably
choosing the gauge fixing parameter.
The singularities
that may appear in the perturbative calculations are studied in details.
Ultraviolet divergences are predicted by the naive power counting,
but it will be shown in Section 3 that they are absent in
the perturbative expansions
of the Green functions.
Still there are spurious singularities, which arise because the
propagators are undamped in the time direction. They are completely
removed with the introduction of a cut off
in the zeroth components of the momenta.
In Section 3, the quantum contributions to the $n-$point
correlation functions are derived
at all orders in perturbation theory.
The one loop case is the most difficult, as
nontrivial cancellations occur among different Feynman diagrams.
To simplify the calculations, a crucial observation is proved, which
drastically
reduces their number.
The total contribution of the
remaining diagrams is shown to vanish
after some algebra. The gluonic $2-$point function
requires some care and it is treated separately.
At two loop, instead, any single Feynman diagram is identically zero.
The reason is that, in order to build such diagrams,
some components of the propagators and of the vertices are required,
which are missing in the Coulomb gauge.
At higher orders, the vanishing of the Feynman diagrams
is proved by induction in the loop number $N$.
Finally, in the Conclusions some open problems and future developments are
discussed.
\section{Chern-Simons Field Theory in the Coulomb Gauge: Feynman Rules and
Regularization}

The C--S action in the Coulomb gauge looks as follows:
\begin{equation}
S_{CS}=S_0+S_{GF}+S_{FP}
\label{action}
\end{equation}
where 

\begin{equation}
S_0=\frac s{4\pi }\int
d^3x\epsilon ^{\mu \nu \rho }\left( \frac
12A_\mu ^a\partial _\nu A_\rho ^a-\frac 16f^{abc}A_\mu ^aA_\nu ^bA_\rho
^c\right)  \label{csaction}
\end{equation}

\begin{equation}
S_{GF}=\frac {is}{8\pi \lambda }\int d^3x\left( \partial
_iA^{a\,i}\right) ^2  \label{gf}
\end{equation}

and 

\begin{equation}
S_{FP}=i\int
d^3x\,\overline{c}^a\partial _i\left( D^i\left[
A\right] c\right) ^a  \label{fp}
\end{equation}

In the above equations $s$ is a dimensionless coupling constant and the
vector fields $A_\mu^a$
represent the gauge potentials. Greek letters $\mu,\nu,\rho,\ldots$
denote space--time indices, while the first latin letters $a,b,c,\ldots=
1,\ldots,N^2-1$ are used for
the color indices of the $SU(n)$ gauge group
with structure constants $f^{abc}$.
The theory is considered on the flat space-time
$\mbox{\bf R}^3$ equipped with the standard euclidean metric
$g_{\mu\nu}=\mbox{\rm  diag}(1,1,1)$.
The total antisymmetric
tensor $\epsilon^{\mu\nu\rho}$ is defined by the convention
$\epsilon^{012}=1$.
Finally,
$$D_\mu^{ab} \left[ A\right] =\partial _\mu \delta ^{ab}-f^{abc}A_\mu ^c$$
is  the covariant derivative and
$\lambda $ is an arbitrary gauge fixing parameter.

In eq. (\ref{action}) the Coulomb gauge constraint is weakly imposed
and the proper Coulomb gauge fixing\footnotemark{}\footnotetext{
From now on, middle latin letters like $i,j,k,\ldots=1,2$ will indicate
space indices.}, given by:
\begin{equation}
\partial _iA^{a\,i}=0  \label{gaugefix}\qquad\qquad\qquad i=1,2
\end{equation}
is recovered setting $\lambda=0$ in eq. (\ref{gf}).

The partition function of the CS field theory
described by eq. (\ref{action}) is: 
\begin{equation}
Z=\int DAD\overline{c}Dce^{iS_{CS}}  \label{partfunct}
\end{equation}
and it is invariant under the BRST transformations listed below:
\begin{eqnarray}
\delta A_\mu ^a &=&\left( D_\mu \left[ A\right] \right) ^a  \label{brst} \\
\delta \overline{c}^a &=&\frac s{4\pi \lambda }\partial _iA^{a\,i}  \nonumber
\\
\delta c^a &=&\frac 12f^{abc}c^bc^c  \nonumber
\end{eqnarray}
%The complex number $i$ ($i^2=-1$) combines with the $i$'s appearing in eqs.
%\ref{gf} and \ref{fp} to give the right signs with which the Faddeev-Popov
%and gauge fixing functionals should appear in the partition function
%in the Euclidean case. On the other side, the pure C--S term $S_0$
%of eq. \ref{csaction} is metric invariant and retains the $i$

From (\ref{action}), it is possible to derive the Feynman rules of C--S
field theory in the Coulomb gauge.
The components of the gauge field
propagator
$G_{\mu \nu }^{ab}(p)$ in the Fourier space
are given by: 
\begin{equation}
G_{jl}^{ab}(p)=
-\delta ^{ab}\frac{4\pi \lambda }s\frac{p_ip_l}{{\mbox{\rm\bf p}}^4}
\label{gjl}
\end{equation}

\begin{equation}
G_{j0}^{ab}(p)=
\delta ^{ab}
\left(
\frac{4\pi }s\epsilon _{0jk}\frac{p^k}{\mbox{\rm\bf p}^2}-
\frac{4\pi \lambda }s
\frac{p_jp_0}{\mbox{\rm\bf p}^4}
\right)
\label{gjo}
\end{equation}

\begin{equation}
G_{0j}^{ab}(p)=
-\delta ^{ab}
\left( 
\frac{4\pi }s
\epsilon _{0jk}
\frac{p^k}{\mbox{\rm\bf p}^2}+
\frac{4\pi \lambda }s\frac{p_0p_j}{\mbox{\rm\bf p}^4}\right) 
\label{goj}
\end{equation}

\begin{equation}
G_{00}^{ab}(p)=
-\delta^{ab}
\frac{4\pi \lambda }s
\frac{p_0^2}{\mbox{\rm\bf p}^4}
\label{goo}
\end{equation}
with $\mbox{\rm\bf p}^2=p_1^2+p_2^2$, while the
ghost propagator $G_{gh}^{ab}(p)$ reads as follows:
\begin{equation}
G_{gh}^{ab}(p)=\frac{\delta ^{ab}}{\mbox{\rm\bf p}^2}  \label{ggh}
\end{equation}
Finally, the three gluon vertex and the ghost-gluon vertex
are respectively given by:
\begin{equation}
V_{\mu _1\mu _2\mu _3}^{a_1a_2a_3}(p,q,r)=-\frac{is}{3!4\pi }(2\pi
)^3f^{a_1a_2a_3}\epsilon ^{\mu _1\mu _2\mu _3}\delta ^{(3)}(p+q+r)
\label{aaa}
\end{equation}
and
\begin{equation}
V_{\mathrm{gh\thinspace }i _1}^{a_1a_2a_3}(p,q,r)=-i(2\pi )^3\left(
q\right) _{i_1}f^{a_1a_2a_3}\delta ^{(3)}(p+q+r)  \label{acc}
\end{equation}
In the above equation we have only given the spatial components of the
ghost-gluon vertex.
From eq. (\ref{fp}), it is in fact easy to realize that
in the Coulomb gauge
its temporal component is zero.

At this point, a regularization should be introduced in order to handle the
singularities that may arise in the computations of the Feynman diagrams.
The potential divergences are
of three kinds.

\begin{enumerate}
\item  Ultraviolet divergences (UV). The naive power counting gives the
following degree of divergence $\omega (G)$ for a given Feynman diagram $G$: 
\begin{equation}
\omega (G)=3-\delta -E_B-\frac{E_G}2  \label{napoco}
\end{equation}
with \footnotemark{}\footnotetext{We use here the same notations of ref.
\cite{itzu}}

\begin{enumerate}
\item  $\delta =$ number of momenta which are not integrated inside the loops

\item  $E_B=$ number of external gluonic legs

\item  $E_G=$ number of external ghost legs
\end{enumerate}

Eq. (\ref{napoco}) shows that UV divergences are possible in the
two and three point functions, both with gluonic or ghost
legs. Moreover, there is also a possible logarithmic divergence in the case
of the four point interaction among two gluons and two ghosts.
In principle, we had to introduce a regularization for these divergences
but in practical calculations this is not necessary.
As a matter of fact, we will see in Section 3 that there are no
UV divergences in the quantum corrections of the Green functions.

\item  Infrared (IR) divergences. 
In the pure C--S field theories \cite{hageno} there are no problems
of infrared divergences.
As a matter of fact, it can be seen from the Feynman rules written above that
the IR behavior of the gluonic propagator is
very mild ($\sim \frac 1{|\mbox{\rm\bf p}|}$). The
potentially more dangerous
IR singularities due to the ghost propagator are screened by the presence of
the external derivative in the ghost--gluon vertex (\ref{acc}).
However, we notice that IR divergences appear
in the interacting case.
For instance, in three dimensional quantum electrodynamics coupled
with a C--S term, the IR divergences have been discussed in refs.
\cite{jao,jat}. 

\item  Spurious divergences.
These singularities appear because the propagators
(\ref{gjl})--(\ref{ggh}) are undamped in the time direction and are
typical of the Coulomb gauge.
To regularize spurious divergences
of this kind, 
it is
sufficient to introduce a cutoff $\Lambda _0>0$ in the domain of integration
over the variable $p_0$: 
\begin{equation}
\int_{-\infty }^\infty dp_0\rightarrow \int_{-\Lambda _0}^{\Lambda _0}dp_0
\label{spureg}
\end{equation}
The physical situation is recovered in the limit $\Lambda _0\rightarrow
\infty $.
\end{enumerate}

\noindent 
As we will see, this regulatization does not cause
ambiguities in the evaluation of the
radiative corrections at any loop order.
In fact, the integrations over the temporal components of the
momenta inside the loops turn out to be trivial and do not interfere
with
the integrations over the spatial components.

\section{Perturbative Analysis}

In this Section we compute the $n-$point
correlation functions of C--S field theories at any loop order.
To this purpose, we choose for simplicity
the proper Coulomb gauge, setting
$\lambda=0$ in eq. (\ref{gf}).
In this
gauge the gluon-gluon propagator has only two nonvanishing components: 
\begin{equation}
G_{j0}(p)=
-G_{0j}(p)=
\delta^{ab}
\frac{4\pi }s
\epsilon _{0jk}\frac{p^k}{
\mbox{\rm\bf p}^2}  \label{gjopcg}
\end{equation}
The presence of $p_0$ remains confined in the vertices
(\ref{aaa})--(\ref{acc})
and it is trivial because it is concentrated in the Dirac $\delta $%
--functions expressing the momentum conservations. As a consequence,
the CS field theory can be considered as a two dimensional model.

First of all we will discuss the one loop calculations.
The following observation greatly reduces the number of diagrams to be
evaluated:

\begin{description}
\item[Observation:] 
Let $G^{(1)}$ be a one particle irreducible (1PI) Feynman diagram containing
only one closed loop. Then all the
internal lines of $G^{(1)}$ are either ghost or gluonic lines.
\end{description}

\begin{figure}
\vspace{1.5truein}
\includegraphics{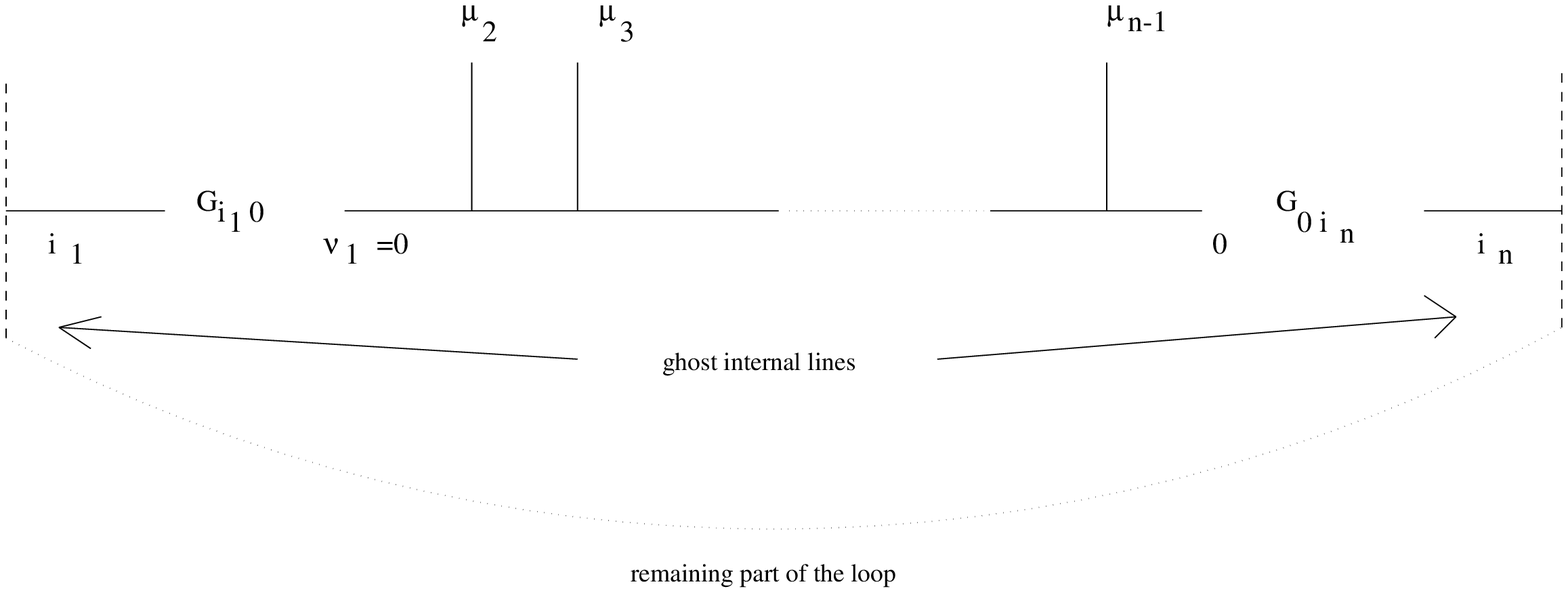}
\vspace{0.25in}
\caption{The figure shows the only possible way in which a tree diagram
$T_{\nu_1\mu_2\ldots\mu_{n-1}\nu_n}$ with
$n$ gluonic legs can be glued to another tree
diagram containing also ghost legs in order to build a
one loop diagram with mixed ghost and gluonic internal lines.}
\label{figtw}
\vspace{1.9truein}
\includegraphics{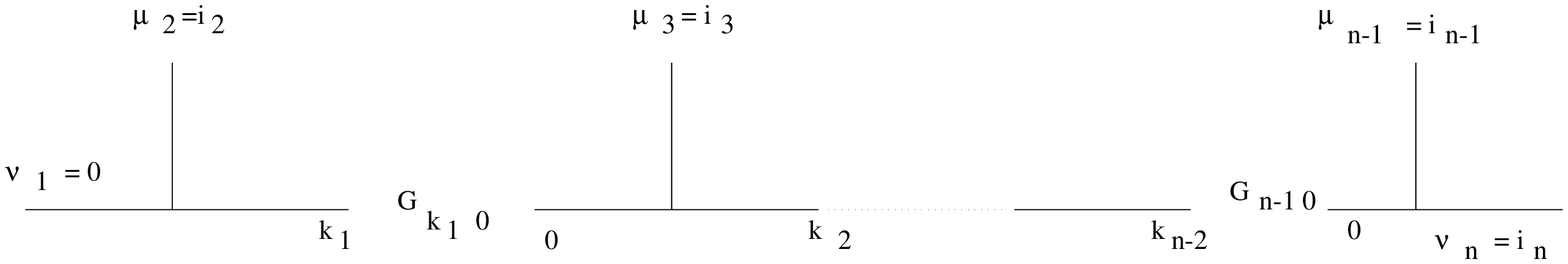}
\vspace{0.25in}
\caption{This  figure shows that in an arbitrary tree diagram
$T_{\nu_1\nu_2\ldots\nu_{n-1}\nu_n}$
constructed in terms of the
gauge fields propagator (\ref{gjopcg})  and the
three gluon vertex (\ref{aaa}), only one component in the
space-time indices $\nu_i$,
$i=1,\ldots,n$, can be temporal.}
\label{figtr}
\end{figure}

To prove the above observation, we notice that the only way to have
a gluonic line preceding or following a ghost line inside a loop 
is to exploit the ghost--gluon  vertex (\ref{acc}).
Thus, if a one loop diagram $G^{(1)}$ with both gluonic and ghost
legs exists, the situation illustrated in fig. \ref{figtw} should occur,
in which at least one gluonic tree diagram
$T_{\nu _1\mu _2...\mu _{n-1}\nu _n}$ is connected to the rest of $G^{(1)}$
by gluing two of its legs, those carrying the indices $\nu_1$ and
$\nu_2$ in the figure, to two ghost--gluon vertices
$V_{\mathrm{gh\thinspace }\nu_1}$ and
$V_{\mathrm{gh\thinspace }\nu_n}$.
At this point, we recall that these vertices have only spatial components
$V_{\mathrm{gh\thinspace }i_1}$ and
$V_{\mathrm{gh\thinspace }i_n}$, $i_1,i_2=1,2$.
As a consequence, since the contractions between gluonic legs are
performed with the propagator (\ref{gjopcg}), it is clear that the necessary
condition for which the whole diagram $G^{(1)}$ does not vanish
is that $\nu_1=\nu_n=0$. On the other side, this is not possible, as
it is shown by fig.
(\ref{figtr}). In fact, because of the presence of an
$\epsilon^{\mu\nu\rho}$ tensor in the gluonic vertex (\ref{aaa}), the most
general gluonic tree diagrams with $n$ legs
$T_{\nu _1\mu _2...\mu _{n-1}\nu _n}$  must have at least
$n-1$ spatial indices in order to be different from zero.
This proves the observation.
An important consequence is that,
at one loop, the only non--vanishing diagrams occur when all the external
legs are gluonic. Hence we have to evaluate only the diagrams describing the
scattering among $n$ gluons. 

This can be done as follows. First of all, we
consider the  diagrams
with internal gluonic lines.
After suitable
redefinitions of the indices and of the momenta, it is possible to see that
their total contribution is given by:
\begin{eqnarray}
V_{i_1...i_n}^{a_1...a_n}
\left(
1;p_1,...,p_n
\right)=
C\left[ -i
\left(2\pi
\right)^3
\right]^n
\frac{n!(n-1)!}2\delta^{(2)}
(\mbox{\rm\bf p}_1+...+
\mbox{\rm\bf p}_n) &&  \label{vone} \\
f^{a_1b_1^{\prime }c_1^{\prime }}f^{a_2b_2^{\prime }b_1^{\prime
}}...f^{a_nc_1^{\prime }b_{n-1}^{\prime }}\int d^2\mbox{\rm\bf q}_1
\frac{\left[
q_1^{i_1}...q_n^{i_n}+q_1^{i_2}\ldots q_j^{i_{j+1}}
\ldots q_{n-1}^{i_n}q_n^{i_1}\right] }{
\mbox{\rm\bf q}%
_1^2...\mbox{\rm\bf q}_n^2} &&  \nonumber
\end{eqnarray}
where $C=\left( 2\Lambda _0\right) ^{2n}$ is a finite constant coming from
the integration over the zeroth components of the momenta and
\begin{equation}
\begin{array}{cccc}
q_2= & q_1+p_1+p_n+p_{n-1}+ & \ldots & +p_3 \\ 
\vdots & \vdots&\ddots&\vdots \\ 
q_j=&q_1+p_1+p_n+p_{n-1}+&\ldots&+p_{j+1} \\ 
\vdots & \vdots&& \\ 
q_n=&q_1+p_1&&
\end{array}
\label{qform}
\end{equation}
for $j=2,\ldots,n-1$.
As it is possible to see from eq. (\ref{vone}),
the only nonvanishing components of $V_{\mu
_1...\mu _n}^{a_1...a_n}\left( 1;p_1,...,p_n\right) $ are those for which $\mu
_1=i_1,$ $\mu _2=i_2,...,\mu _n=i_n$, i. e. all tensor indices
$\mu_1,\ldots,\mu_n$
are
spatial.

The case of the Feynman diagrams containing ghost internal lines
is more complicated. After some work, it is possible to
distinguish two different contributions to the Green functions
with $n$ gluonic legs:
\begin{eqnarray}
V_{i_1...i_n}^{a_1...a_n}\left( 2a;p_1,...,p_n\right)=-C\left[ -i\left(
2\pi \right) ^3\right]^n\frac{n!(n-1)!}2
&&  	\nonumber \\
\delta^{(2)}(\mbox{\rm\bf p}_1+...%
\mbox{\rm\bf p}_n)
f^{a_1b_1^{\prime }c_1^{\prime }}f^{a_2b_2^{\prime }b_1^{\prime
}}...f^{a_nc_1^{\prime }b_{n-1}^{\prime }}\int d^2\mbox{\rm\bf q}_1\frac{%
q_1^{i_1}...q_n^{i_n}}{\mbox{\rm\bf q}_1^2...
\mbox{\rm\bf q}_n^2}\label{vtwoa}&&
\end{eqnarray}
and 
\begin{eqnarray}
V_{i_1...i_n}^{a_1...a_n}\left( 2b;p_1,...,p_n\right) =C\left( -1\right)
^{n-1}\left[ -i\left( 2\pi \right) ^3\right] ^n\frac{n!(n-1)!}2 &&
\nonumber \\
\delta ^{(2)}(\mbox{\rm\bf p}_1+...\mbox{\rm\bf p}_n)
f^{a_1b_1^{\prime }c_1^{\prime
}}f^{a_2b_2^{\prime }b_1^{\prime }}...f^{a_nc_1^{\prime }b_{n-1}^{\prime
}}\int d^2\mbox{\rm\bf q}'_1\frac{(q_1')^{i_1}...(q_n')^{i_n}}
{(\mbox{\rm\bf q}_1')^2...
(\mbox{\rm\bf q}_n')^2} &&
\label{vtwob}
\end{eqnarray}
where the constant $C$ is the result of the integration over the
zeroth components
of the momenta and it is the same of eq. (\ref{vone}). 
Apart from an overall sign, eqs. (\ref{vtwoa}) and (\ref{vtwob}) differ
also by the
definitions of the momenta. In (\ref{vtwoa}) the variables $q_2,...,q_n$ are in
fact given by  eq. (\ref{qform}). In eq. (\ref{vtwob}) we have instead: 
\begin{equation}
\begin{array}{cccc}
q_2'=&q_1'+p_1&& \\ 
\vdots & \vdots&& \\ 
q_j'=&q_1'+p_1+&\ldots&+p_{j-1} \\ 
\vdots & \vdots&\ddots&\vdots \\  
q_n'=&q_1'+p_1+&\dots&+p_{n-1}
\end{array}
\label{qformb}
\end{equation}
for $j=2,\ldots,n-1$.

To compare eq. (\ref{vtwob}) with  (\ref{vone}) and (\ref{vtwoa}) we
perform the change of variables
\begin{equation}
q_1=-q_1'-p_1
\label{shift}
\end{equation}
in eq. (\ref{vtwob}).
Exploiting eq. (\ref{shift}) and
the relation $p_1+...+p_n=0$, we obtain:
\[
V_{i_1...i_n}^{a_1...a_n}\left( 2b;p_1,...,p_n\right) = 
-C[-i(2\pi)^3]^n
\frac{n!(n-1)!}2f^{a_1b_1^{\prime
}c_1^{\prime }}f^{a_2b_2^{\prime }b_1^{\prime }}...f^{a_nc_1^{\prime
}b_{n-1}^{\prime }}
\]
\begin{equation}
\delta ^{(2)}(\mbox{\rm\bf p}_1+...+\mbox{\rm\bf p}_n) 
\int d^2\mbox{\rm\bf q}_1
\frac{q_n^{i_1}q_1^{i_2}\ldots q_j^{i_{j+1}}\ldots q_{n-1}^{i_n}}
{(\mbox{\rm\bf q}_1')^2...
(\mbox{\rm\bf q}_n')^2}\label{finvtb}
\end{equation}
where the variables $q_2,\ldots,q_n$ are now defined as in eq. (\ref{qform}).
At this point we can sum 
eqs. (\ref{vone}),
(\ref{vtwoa}) and (\ref{finvtb}) together.
It is easy to realize that the total result is zero, i. e.:

\begin{equation}
V_{i_1...i_n}^{a_1...a_n}\left( 1;p_1,...,p_n\right)
+V_{i_1...i_n}^{a_1...a_n}\left( 2a;p_1,...,p_n\right)
+V_{i_1...i_n}^{a_1...a_n}\left( 2b;p_1,...,p_n\right) =0 \label{finres}
\end{equation}

Still, it is not possible to conclude from eq. (\ref{finres}) that
there are no radiative corrections at one loop in
C--S field theory.
Let us remember in fact that
eq. (\ref{finres})  has been obtained from eq. (\ref{vtwob}) after performing
the shift of variables
(\ref{shift}). This could be dangerous
if there are unregulated
divergences.
However, it is not difficult to verify that
each of the integrals appearing
in the right hand sides of eqs. (\ref{vone}), (\ref{vtwoa})
and (\ref{vtwob}) is IR and UV finite for $n\ge 3$.
Only the case $n=2$ needs some more care.
Summing together eqs. (\ref{vone}), (\ref{vtwoa}) and (\ref{finvtb})
for $n=2$, we obtain the following result:
%\[
%V_{ij}^{ab}\left( 1;p_1,p_2\right) =-\left( 2\pi \right) ^6\left(
%2\Lambda _0\right) ^2N\delta ^{ab}\delta ^{(2)}(\mbox{\rm\bf p}_1+
%\mbox{\rm\bf p}%
%_2)
%\]
%\[
%\int d^2\mbox{\rm\bf q}
%\frac{\left[ q_{i}\left( q+p_1\right)_{j}+\left(
%q+p_1\right)_{i}q_{j}\right] }{\mbox{\rm\bf q}^2\left(
%\mbox{\rm\bf q}+%
%\mbox{\rm\bf p}_1\right) ^2} 
%\]
%\[
%V_{ij}^{ab}\left( 2a;p_1,p_2\right) =\left( 2\pi \right) ^6\left(
%2\Lambda _0\right) ^2N\delta ^{ab}\delta ^{(2)}(\mbox{\rm\bf p}_1+
%\mbox{\rm\bf p}%
%_2)\int d^2\mbox{\rm\bf q}
%\frac{q_{i}\left( q+p_1\right)_{j}}{\mbox{\rm\bf q}%
%^2\left( \mbox{\rm\bf q}+\mbox{\rm\bf p}_1\right) ^2} 
%\]
%\[
%V_{ij}^{ab}\left( 2b;p_1,p_2\right) =\left( 2\pi \right) ^6\left(
%2\Lambda _0\right) ^2N\delta ^{ab}\delta ^{(2)}(\mbox{\rm\bf p}_1+
%\mbox{\rm\bf p}%
%_2)\int d^2\mbox{\rm\bf q}\frac{q_{i}\left( q+p_1\right)_{j}}
%{\mbox{\rm\bf q}%
%5^2\left( \mbox{\rm\bf q}+\mbox{\rm\bf p}_1\right) ^2} 
%\]
%
\[
V_{ij}^{ab}\left( 1;p_1,p_2\right) +V_{ij}^{ab}\left(
2a;p_1,p_2\right) +V_{ij}^{ab}\left( 2b;p_1,p_2\right) = 
\]
\begin{equation}
\left( 2\pi \right) ^6\left( 2\Lambda _0\right) ^2N\delta ^{ab}\delta
^{(2)}(\mbox{\rm\bf p}_1+\mbox{\rm\bf p}_2)
\int d^2\mbox{\rm\bf q}\frac{\left[ q_{i}
(p_1)_{j}-q_{j}(p_1) _{i}
\right] }
{\mbox{\rm\bf q}^2\left( \mbox{\rm\bf q}+\mbox{\rm\bf p}_1\right) ^2}
\label{cru}
\end{equation}
where we have put $q_1'=q_1=q$.
As we see, the integrand appearing
in the rhs of (\ref{cru})
is both IR and UV
finite.  Moreover, 
a simple computation shows that 
the integral over $\mbox{\rm\bf q}$
is zero without the need of the shift (\ref{shift}).
As a consequence, there are no
contributions to the Green functions at one loop.

Now we are ready to consider the higher order corrections.
At two loop, a general Feynman diagram $G^{(2)}$ can be obtained
contracting two legs of a tree diagram $G^{(0)}$ with
two legs of a one loop diagram $G^{(1)}$.
As previously seen, the latter have only gluonic
legs and their tensorial indices are all spatial.
Consequently, in order to perform the contractions by means of the propagator
(\ref{gjopcg}),  there should exist one component of 
$G^{(0)}$ with at least two temporal indices,
but this is impossible. To convince
oneself of this fact, it
is sufficient to look at fig. (\ref{figtr}) and related comments.
The situation does not improve
if we build $G^{(0)}$  exploiting also the ghost-gluon vertex
(\ref{acc}), because it has no temporal component.
As a consequence, all the Feynman graphs vanish identically at two loop order.
Let us notice that it is possible to verify their vanishing
directly, since
the number of two loop diagrams
is relatively small in the Coulomb gauge and one has just to contract the
space-time indices without performing the integrations over the internal
momenta.
However, this
procedure is rather long and will not be reported here.

Coming to the higher order computations, we notice that
a diagram with $N+1$ loops $G^{(N+1)}$
has at least
one subdiagram $G^{(N)}$
containing $N-$loops.
Supposing that $G^{(N)}$ is identically equal to
zero because it cannot be constructed with the
Feynman rules (\ref{ggh})--(\ref{acc}) and (\ref{gjopcg}), also $G^{(N+1)}$
must be zero.
As we have seen above, there are no Feynman diagrams for $N=2$.
This is enough to prove by induction that
the C--S field theories have  no radiative corrections
in the Coulomb gauge
for any value of $N$.

\section{Conclusions}

In this paper we have proved with explicit computations that the C--S field
theories do not have quantum corrections in the Coulomb gauge.
At two loop order and beyond, this is a trivial consequence of the
fact that it is impossible to construct nonzero Feynman diagrams
starting from the vertices and propagators given in eqs.
(\ref{ggh})--(\ref{acc}) and (\ref{gjopcg}).
At one loop, instead, nontrivial cancellations occur
between the different diagrams.
We have also seen that the perturbative
expansion of the Green functions is not affected by
UV or IR divergences.
Only the spurious singularities are present, which are related to the fact
that the propagators are undamped in the time direction.
They are similar to the singularities observed in the four dimensional
Yang--Mills field theories
\cite{taylor}, but in the C--S case appear in a milder form.
In fact, after the regularizarion (\ref{spureg}), their
contribution at any loop order reduces to
a factor in the radiative
corrections and does not influence
the remaining calculations.
Therefore, the results obtained here are regularization
independent.
Moreover, the vanishing of the quantum contributions described in Section 3
is a peculiarity of the Coulomb gauge that does not strictly depend from the
fact that the C--S field theories are topological.
An analogous situation occurs in the light cone gauge in the presence of a
boundary. In that case, radiative corrections arise due to the interactions
of the fields with the boundary, but each Feynman diagram corresponding to
these interactions vanishes identically \cite{empi}.

In summary, our study indicates that the Coulomb gauge is a
convenient and reliable gauge fixing, especially in the
perturbative applications of C-S field theory.
Let us remember that, despite of the fact that the theory does non
contain degrees of freedom, the perturbative calculations
play a relevant role, for instance in the computations of knot invariants
\cite{alr}, \cite{witten}--\cite{axelrod}.
Contrary to what happens using the covariant gauges,
where it becomes more and
more difficult to evaluate the radiative corrections
as the loop number increases \cite{alr,gmm,chaichen}, 
in the Coulomb gauge
only the tree level contributions to the Green
functions survive. This feature is
particularly useful in the
case of non-flat manifolds, where the momentum representation does not exist.
For instance, Feynman rules analogous to those given in
eqs. (\ref{gjl})--(\ref{acc})
have been derived also on the compact Riemann surfaces
\cite{ffunp}.
In the future, besides the applications in knot theory, we plan to extend
our work also to C--S field theories with non-compact gauge group, in order
to include also the theory of quantum gravity in $2+1$ dimensions.
Moreover, most of the pathologies that seem to afflict the four dimensional
gauge field theories, like spurious
and infrared divergences, are also present in the C--S field
theories,
but in a milder form. As a consequence, the latter can be considered
as a good laboratory in order to study their possible remedies.
For example, it would be interesting
to apply to the Yang--Mills case
the regularization (\ref{spureg}) introduced here for the spurious
singularities.
Let us
notice that a different regularization
has been recently proposed in \cite{leiwil}.
Finally, the present analysis is limited to the pure
C--S field theories and more investigations
are necessary for the interacting case.
Until now, only the models based on abelian C--S field theory
have been studied in details, in particular
the
so-called Maxwell-Chern-Simons field theory, whose consistency
 in the Coulomb gauge
has been checked with several tests \cite{devone}.
%A physical application of our results,
%which is currently under consideration, is the
%investigation of the statistics of fermionic and bosonic matter
%fields interacting with nonabelian C--S theories at high temperatures
%\cite{higtemp}. Other interesting applications are $(2+1)$
%quantum gravity and the calculation of the new link invariants
%from C--S field theories quantized
%on Riemann surfaces, whose existence has been formally shown
%in \cite{cotta}. In these latter two cases,
%the possibility offered by the Coulomb gauge of performing explicit
%calculations also on non--flat space--times \cite{ffprd}
%can be exploited.


\begin{thebibliography}{99}
\bibitem{jao} R. Jackiw and S. Templeton, {\it Phys. Rev.} {\bf D23} (1981),
2291; S. Deser, R. Jackiw and S. Templeton, {\it Phys. Rev. Lett.}
{\bf 48} (1983), 975;
J. Schonfeld, {\it Nucl. Phys.} {\bf B185} (1981), 157.
\bibitem{hageno}C. R. Hagen, {\it Ann. Phys.} (NY) {\bf 157} (1984), 342.
\bibitem{csapps} G. Moore and N. Seiberg, {\it Phys. Lett.} {\bf B220} (1989),
422; J. Fr\"ohlich and C. King, {\it Comm. Math. Phys.} {\bf 126}
(1989), 167; J. M. F. Labastida and A. V. Ramallo, {\it Phys. Lett.}
{\bf 238B} (1989), 214; M. Bos and V. P. Nair,
{\it Int. Jour. Mod. Phys.} {\bf A5} (1990), 989;
%conformal field theories
W. Chen, G. W. Semenoff and Y. S. Wu, {\it Mod. Phys. Lett.} {\bf A5}
(1990), 1833 {\it Phys. Rev.} {\bf D46} (1992), 5521;
E. Guadagnini, M. Martellini and M. Mintchev, {\it Nucl. Phys.}
{\bf B336} (1990), 581; %calcoli perturbativi degli invarianti di link
G. W. Semenoff, {\it Phys. Rev. Lett} {\bf 61} (1988), 517; E. Fradkin {\it
Phys. Rev. Lett.} {\bf 63} (1989), 322; M. L\"uscher, {\it Nucl. Phys.}
{\bf B326} (1989), 557; %t-J models
E. Witten, {\it Nucl. Phys.} {\bf B311} (1988), 46; %(2+1) Quantum Gravity
Y. H. Chen, F. Wilczek, E. Witten and B. I. Halperin, {\it Int. Jour. Mod.
Phys.} {\bf B3} (1989), 1001;
\bibitem{ienkur}
R. Iengo and K. Lechner, {\it Phys. Rep.} {\bf 213} (1992), 179.
% anyons models in Tc Suco and in FQHE
\bibitem{ffprd} F. Ferrari, {\it Phys. Rev.} {\bf D50} (1994), 7578.
\bibitem{hagent} C. R. Hagen, {\it Phys. Rev.} {\bf D31} (1985),
2135.
\bibitem{jat} S. Deser, R. Jackiw and S. Templeton,
{\it Ann. Phys.} (N. Y.) {\bf 140} (1984), 372.
\bibitem{vo}  A. Desnieres de Veigy and S. Ouvry, {\it Phys. Lett.} {\bf 387B%
} (1993), 91; D. Bak and O. Bergman, {\it Phys.
Rev.} {\bf D51} (1995), 1994; O. Bergman and G. Lozano,
{\it Ann. Phys.} (NY) {\bf 229} (1994), 229.
\bibitem{devone} F. P. Devecchi, M. Fleck, H. O. Girotti, M. Gomes and A. J. da Silva,
{\it Ann. Phys.} {\bf 242} (1995), 275.
\bibitem{devtwo} A. Foerster and H. O. Girotti, {\it Statistical transmutations
in $2+1$ dimensions}, in J. J. Giambiagi Festschrift, eds. H. Falomir, R. E.
Gamboa Saravi, P. Leal Ferreira and F. A. Schaposnik (World Scientific,
Singapore, 1990), p. 161; {\it Phys. Lett.} {\bf B230} (1989), 83;
{\it Nucl. Phys.} {\bf B342} (1990), 680.
\bibitem{bcv} A. Bellini, M. Ciafaloni and P. Valtancoli, {\it Nucl. Phys.}
{\bf B454} (1995), 449; {\bf B462} (1996), 453.
\bibitem{taylor} P. J. Doust and J. C. Taylor, {\it Phys. Lett.}
{\bf 197B} (1987), 232; P. J. Doust, {\it Ann. Phys.} (N. Y.) {\bf 177} (1987),
169.
\bibitem{chetsa} H. Cheng and E. C. Tsai, {\it Phys. Rev. Lett.}
{\bf 57} (1986), 511. 
\bibitem{leibbrandt} G. Leibbrandt, {\it Noncovariant Gauges}, World
Scientific, Singapore, 1994.
\bibitem{leiwil}
G. Leibbrandt and J. Williams,
{\it Nucl. Phys. } {\bf B475} (1996), 469 and references therein.
\bibitem{cscgform} K. Haller and E. L. Lombridas, {\it Ann. Phys.} {\bf 246} (1996), 1;
M.-I. Park and Y.-J. Park, {\it Phys. Rev.} {\bf D50} (1994),
7584.  
\bibitem{frig} F. Ferrari and I. Lazzizzera, {\it Dirac Quantization
of the Chern--Simons Field Theory in the Coulomb Gauge},
Preprint UTF 386/96, PAR-LPTHE 96-38, to be published in {\it Phys. Lett.}
{\bf B}.
\bibitem{csformal} A. Blasi and R. Collina, {\it Nucl. Phys.} {\bf B354}
(1990), 472; F. Delduc, C. Lucchesi, O. Piguet and S. P. Sorella,
{\it Nucl. Phys.} {\bf B346} (1990), 313.
\bibitem{ss} O. Piguet and S. P. Sorella, {\it Algebraic Renormalization},
Springer Verlag 1995.
\bibitem{shift} G. Giavarini, C. P. Martin and F. Ruiz Ruiz,
{\it Phys. Lett.} {\bf B332} (1994), 345; {\it Phys. Lett.} {\bf B314}
(1993), 328;
{\it Nucl. Phys.}
{\bf B381} (1992), 222; C. P. Martin, {\it Phys. Lett.} {\bf B241} (1990),
513; M. Asorey, F. Falceto, J. L. Lopez and G. Luz\'on, {\it Phys. Rev.}
{\bf D49} (1994), 5377; M. Asorey and F. Falceto, {\it Phys. Lett.}
{\bf B241} (1990), 31.
\bibitem{alr} L. Alvarez--Gaum\'e, J. M. F. Labastida and A. V. Ramallo,
{\it Nucl. Phys.} {\bf B334} (1990), 103;
\bibitem{itzu} C. Itzykson and J.-B. Zuber, {\it Quantum Field Theory},
McGraw--Hill, Singapore 1980.
\bibitem{empi} S. Emery and O. Piguet, {\it Helv. Phys. Acta} {\bf 64}
(1991), 1256.
\bibitem{witten} E. Witten, {\it Comm. Math. Phys.} {\bf 121} (1989), 351. 
\bibitem{gmm} E. Guadagnini, M. Martellini and M. Mintchev,
{\it Phys. Lett.} {\bf B227} (1989), 111.
\bibitem{cotta} P. Cotta Ramusino, E. Guadagnini, M. Martellini and
M. Mintchev, {\it Nucl. Phys.} {\bf B330} (1990), 557.
\bibitem{axelrod} S. Axelrod and I. M. Singer, in {\it Proceedings
of the XXth International Conference on Differential Geometrical
Methods in Theoretical Physics}, New York 1991,  
edited by S. Catto and A. Rocha (World Scientific, Singapore, 1992).
\bibitem{chaichen} M. Chaichian and W. F. Shen, {\it Two loop
Finiteness of Chern--Simons Field Theory in Background Field
Method}, hep-th/9607208.
\bibitem{ffunp} F. Ferrari, {\it On the Quantization of the Chern--Simons
Field Theory on Curved Space--Times: the Coulomb Gauge Approach},
Preprint LMU-TPW 96-5, hep-th/9303117. 
%\bibitem{hagt} C. R. Hagen, {\it Phys. Rev.} {\bf D31} (1985), 331.




%D. Birmingham, M. Rakowsky and G. Thompson, {\it Phys. Lett.}
%{\bf B251} (1990), 121;

%\bibitem{higtemp} G. Dunne, R. Jackiw, S.-Y. Pi and C. A. Trugenberger,
%{\it Phys. Rev.} D {\bf 43} (1991), 1332; G. Dunne, {\it Comm. Math. Phys.}
%{\bf 150} (1993, 519.

\end{thebibliography}
\end{document}